\documentclass[12pt]{article}
\usepackage{scicite}
\usepackage{times}

\usepackage{graphicx}
\usepackage{epsfig}
\usepackage{amsmath}
\usepackage{amssymb}
\newenvironment{sciabstract}{%
\begin{quote} \bf}
{\end{quote}}

\newcounter{lastnote}

\title{Room temperature spin-ice physics in\\ cadmium cyanide}

\author
{Chloe S. Coates,$^1$ Mia Baise,$^2$ Arkadiy Simonov,$^1$ Joshua W. Makepeace,$^1$\\
Andrew G. Seel,$^{1,3}$ Ronald I. Smith,$^4$ Helen Y. Playford,$^4$ David A. Keen,$^4$\\
Ren{\'e}e Siegel,$^5$ Adrian Schmutzler,$^5$ J{\"u}rgen Senker,$^5$ Ben Slater,$^{2\ast}$\\
 and Andrew L. Goodwin$^{1\ast}$\\
\\
\normalsize{$^{1}$Department of Chemistry, University of Oxford, Inorganic Chemistry Laboratory,}\\
\normalsize{South Parks Road, Oxford OX1 3QR, U.K.}\\
\normalsize{$^{2}$Department of Chemistry, University College London, Gower Street,}\\
\normalsize{London WC1E 6BT, U.K.}\\
\normalsize{$^{3}$Department of Physics and Astronomy, University College London,}\\
\normalsize{Gower Street, London WC1E 6BT, U.K.}\\
\normalsize{$^{4}$ISIS Facility, Rutherford Appleton Laboratory, Harwell Campus, Didcot,}\\
\normalsize{Oxfordshire OX11 0QX, U.K.}\\
\normalsize{$^{5}$Anorganische Chemie III, University of Bayreuth, Universitätsstr. 30,}\\
\normalsize{95447 Bayreuth, Germany}\\
\\
\normalsize{E-mail: b.slater@ucl.ac.uk, andrew.goodwin@chem.ox.ac.uk.}
}
\date{}

\begin{document} 

\baselineskip24pt

\maketitle 

\begin{sciabstract}

Spin-ices are frustrated magnets that support a particularly rich variety of emergent physics. Typically, it is the interplay of magnetic dipole interactions, spin anisotropy, and geometric frustration on the pyrochlore lattice that drives spin-ice formation. The relevant physics occurs at temperatures commensurate with the magnetic interaction strength, which for most systems is 1--5\,K. This low energy scale poses severe challenges for experimental studies of spin-ices and the practical exploitation of their unusual properties. Here, we show that non-magnetic cadmium cyanide (Cd(CN)$_2$) exhibits analogous behaviour to magnetic spin-ices, but does so on a temperature scale that is nearly two orders of magnitude greater. The electric dipole moments of cyanide ions in Cd(CN)$_2$ assume the role of magnetic pseudospins, with the difference in energy scale reflecting the increased strength of electric \emph{vs} magnetic dipolar interactions. As a result, spin-ice physics influences the structural behaviour of Cd(CN)$_2$ even at room temperature.

\end{sciabstract}

The pyrochlore lattice of vertex-sharing tetrahedra is a recurring motif in many classes of geometrically frustrated materials \cite{Anderson_1956,Bramwell_1998,Moessner_2006,Trump_2018}. Amongst these, systems for which each vertex is associated with an Ising variable ($e=\pm1$, say) and which obey a constant-sum rule on each tetrahedron ($\sum_ie_i=0$) form the particularly intriguing family of ``ices'' \cite{Henley_2010}. (Cubic) water ice \cite{Konig_1943}  and spin-ice Dy$_2$Ti$_2$O$_7$ \cite{Bramwell_2001} are two examples [Fig.~1A,B]; many others are known \cite{McQueen_2008,Melot_2009,McClarty_2014,Thygesen_2017}. Common to all ice-like states is a huge configurational degeneracy---reflected in the Pauling entropy \cite{Pauling_1935,Ramirez_1999}---that in principle allows these systems to be exploited in data storage and manipulation \cite{Nisoli_2013,Wang_2016}. Moreover, the constant-sum rule ($\equiv$ ``ice rule'' \cite{Bernal_1933}) leads to an effective gauge field that can in turn drive a variety of remarkable physics \cite{Henley_2010,Powell_2011,Benton_2012,Fennell_2019}. For example, violations of this rule (excitations of the gauge field) behave as emergent quasiparticles that interact with one another \emph{via} an effective Coulomb potential \cite{Henley_2010,Morris_2009}. These particles represent a fractionalisation of the underlying Ising variable, such that in the spin-ices they behave as magnetic monopoles (\emph{i.e.}\ fractionalised magnetic dipoles) \cite{Castelnovo_2008,Ladak_2010,Castelnovo_2012}. The manipulation of monopoles with external fields is thought to be a promising avenue for developing novel spintronic devices \cite{Khomskii_2012}.

Of particular practical importance in seeking to apply this unusual physics is the energy scale that governs a given ice-like phase. How difficult is it to invert Ising states? And how strictly are ice rules obeyed? In water ice the energies are simply too high; hydrogen-bond inversion is usually sluggish and ice rule violations are exceedingly rare ($\sim$1 ppm at 260\,K) \cite{Bjerrum_1952,Hobbs_1974,Cowin_1999,deKoning_2008}. By contrast, spin-ices are dynamic to very low temperatures ($<1$\,K), but the energy cost of defect formation is comparably small \cite{Matsuhira_2011,Morris_2009}. Hence spin-ice physics is constrained to the single-Kelvin regime.

Motivated by the potential impact of identifying ice-like phases with more advantageous energetics, we study the molecular framework material cadmium cyanide, Cd(CN)$_2$. Its anticuprite structure contains cyanide ions situated on the vertices of a pair of interpenetrating pyrochlore lattices \cite{Shugam_1945} [Fig.~1C]. At ambient temperature, the crystal symmetry is $Pn\bar3m$ and the system is isostructural with high-pressure proton-disordered ice-VII \cite{Kuhs_1984}: the O position is occupied by Cd and the (average) H position by CN, with head-to-tail orientational disorder \cite{Nishikiori_1991}. Solid-state NMR and single-crystal X-ray diffuse scattering measurements, together with density-functional theory (DFT) calculations, have collectively identified Cd(CN)$_2$ as a candidate ice \cite{Nishikiori_1990,Nishikiori_1991,Ding_2008,Fairbank_2012}. The orientation of each individual CN$^-$ ion acts as an Ising variable, and the constant-sum rule reflects a preference for each Cd to bind two C and two N atoms \cite{Fairbank_2012}, evoking the ice rules.

What is entirely unknown is whether CN$^-$ flipping is possible in Cd(CN)$_2$, and hence whether the system is capable---even in principle---of exhibiting spin-ice physics. In fact, our collective understanding of the lattice dynamics of this system is conspicuously poor. For example, on cooling to $\sim$130\,K the material exhibits a displacive phase transition that is not only uncharacterised \cite{Fairbank_2012}, but is entirely unexpected: DFT calculations find no evidence of lattice instabilities in the parent phase \cite{Ding_2008,Zwanziger_2007}. From an experimental viewpoint, there are a number of reasons why structural and dynamical studies of Cd(CN)$_2$ are particularly complicated: one is the inability for X-ray scattering measurements to distinguish CN$^-$ orientations, especially in the presence of electron-rich Cd$^{2+}$ ions \cite{Hoskins_1990}; a second is the extreme sensitivity of Cd(CN)$_2$ to damage from X-ray beams, which affects the reproducibility of phase-transition and thermal expansion behaviour \cite{Coates_2018}; and a third is the (in)famously high neutron absorption cross-section of natural-abundance Cd \cite{DMellow_2007}, complicating both elastic and inelastic neutron scattering measurements.

Using a recently devised synthetic route to isotopically enriched Cd(CN)$_2$ \cite{Coates_2018}, we prepared a polycrystalline sample of $^{114}$Cd(CN)$_2$ suitable for neutron scattering measurements. This sample has allowed us for the first time to characterise the structure of Cd(CN)$_2$ and its temperature dependence without the complications of X-ray sensitivity. Our results are shown in Fig.~2. On cooling from room temperature, the $Pn\bar3m$ cubic unit cell of ambient-phase Cd(CN)$_2$ expands (hence negative thermal expansion, NTE \cite{Goodwin_2005}), until at 130\,K a structural phase transition occurs. We find the low-temperature phase to have tetragonal $I4_1/amd$ symmetry and to be isostructural to hydrogen-ordered ice-VIII \cite{Kuhs_1984}. Specifically, its crystal symmetry now allows for long-range CN$^-$ orientational order, and we do indeed find progressive ordering on cooling---evidenced by a systematic change in scattering density at the two crystallographically-distinct C/N sites---until an apparent orientational glass transition at $\sim$80\,K. No further structural transitions were observed to 10\,K \cite{notenotcaseforxrays}. The non-equilibrium nature of the 80\,K transition was verified by repeated heating/cooling cycles, which showed subtle but sensible history dependencies. The crucial point of course is that we observe the emergence of CN$^-$ orientational order; this is possible \emph{only if} CN$^-$ flips are thermally accessible, which is clearly the case for $T>80$\,K. So our data establish Cd(CN)$_2$ as a genuine candidate for spin-ice physics. 

We proceed to determine whether a suitable spin-ice Hamiltonian can succeed in capturing the key behaviour of Cd(CN)$_2$, and in turn be tested against further experimental observations. Our starting point is the anisotropic Heisenberg model first proposed in Ref.~\citenum{Champion_2002}:
\begin{equation}\label{ham1}
\mathcal H=-J_{\rm{eff}}\sum_{i,j}\mathbf S_i\cdot\mathbf S_j-\Delta\sum_i S_{i\parallel}^2,
\end{equation}
where the pairwise sum is over nearest-neighbour spin sites $i,j$. This model develops spin-ice behaviour at $T\sim O(J_{\rm{eff}})$ for large $\Delta$ (strong single-ion anisotropy) and for ferromagnetic nearest-neighbour (effective) exchange interactions $J_{\rm{eff}}>0$. Its ground state is ordered for all finite $\Delta$, with the ordering transition temperature suppressed as the Ising limit is approached ($\Delta\rightarrow\infty$) \cite{Champion_2002}. In our mapping, the unit vectors $\mathbf S_i$ represent CN$^-$ dipole orientations and behave as classical Heisenberg pseudospins. We use continuous rather than Ising variables because thermal fluctuations mean the Cd--CN--Cd linkages will not be entirely linear at the temperature ranges we have probed experimentally \cite{Chapman_2005}. Ising-like anisotropy is introduced by the second term in Eq.~(\ref{ham1}): $S_{i\parallel}$ denotes the projection of CN$^-$ orientation vector $\mathbf S_i$ onto the unit vector spanning its two connected Cd centres. The parameter $\Delta$ arises from the local crystal field at the CN$^-$ site and describes the barrier height to CN$^-$ flipping. The exchange term in Eq.~(\ref{ham1}) will have two main contributions for Cd(CN)$_2$: one is a chemical bonding or covalency term---we denote this component by $J$---and the second arises from dipole--dipole interactions. Since electric and magnetic dipolar interactions have the same functional form, we can make use of the established results for the pyrochlore lattice \cite{denHertog_2000,Melko_2004} that long-range dipolar interactions can be effectively truncated at nearest neighbour and are described by an effective exchange term $-5D\sum_{i,j}\mathbf S_i\cdot\mathbf S_j$. Here, $D$ is the electric dipole interaction strength. Taking into account the dipolar coupling between the two interpenetrating pyrochlore lattices in Cd(CN)$_2$ we arrive at our model Hamiltonian:
\begin{equation}\label{ham2}
\mathcal H=-J_{\rm{eff}}\sum_{i,j}\mathbf S_i\cdot\mathbf S_j+D\sum_{i,j^\prime}\mathbf S_i\cdot\mathbf S_{j^\prime}-\Delta\sum_i S_{i\parallel}^2,
\end{equation}
where $J_{\rm{eff}}=J+5D$ and the prime notation indicates nearest neighbours from alternate lattices [Fig.~3A]. The dipolar coupling coefficients $D$ that enter the first two terms of Eq.~(\ref{ham2}) are identical because nearest-neighbour CN pairs lie at equivalent distances ($=a/\sqrt{2}$) whether they belong to the same or to different pyrochlore lattices; the difference in prefactors ($-$5 and $+$1, respectively) is a geometric result.

Quantum chemical (QC) calculations allow us to estimate the magnitude of the various parameters in Eq.~(\ref{ham2}) relevant to Cd(CN)$_2$. We determined the QC energies of a range of small single- and double-network Cd(CN)$_2$ unit cells with different CN$^-$ orientation decorations, testing for consistency by employing a variety of density functionals. The energies of these configurations can be interpreted in terms of $J_{\rm{eff}}$ and $D$, giving $145<J_{\rm{eff}}<191$\,K and $15<D<144$\,K [Fig.~3B,C]; in the SI we elaborate on why it is so difficult computationally to capture $D$ for this particular system. The flipping barrier height $\Delta$ was determined using nudged elastic band calculations \cite{neb_ref} to trace the energy profile during a single CN$^-$ flip. The lowest-barrier mechanism always involves a C-bridged Cd--(CN)--Cd transition state and a barrier height $10\,500<\Delta<13\,000$\,K [Fig.~3D]. As independent checks on these values we note the following. Since the CN$^-$ ion carries a dipole moment of ~1.0\,D, we would estimate $D\simeq85$\,K from geometric arguments. The Cd coordination distributions measured by NMR at 298\,K suggest $J_{\rm{eff}}\simeq140$--$170$\,K \cite{Nishikiori_1990}. And the flipping barrier heights for alkali metal cyanides of $\sim$1500\,K \cite{Sethna_1985} give an extreme lower bound for $\Delta$. So there is good collective consistency with our QC results. We will come to show that the key physics of Eq.~(\ref{ham2}) are actually invariant across the entire ensemble of QC parameters, but we take hereafter the PBE+D3 \cite{grimme_d3} result ($J_{\rm{eff}}=191$\,K, $D=93$\,K, $\Delta=12\,800$\,K) as representative. Remarkably, the \emph{relative} energy scales of these different terms mirror those in spin-ice Dy$_2$Ti$_2$O$_7$, for which $J_{\rm{eff}}=3.3$\,K, $D=1.41$\,K, and $\Delta\sim200$\,K \cite{denHertog_2000,Tomasello_2015,noterefactor}; we have $D/J_{\rm{eff}}=0.43\ (0.49)$ and $\Delta/J_{\rm{eff}}=61\ (67)$ for Dy$_2$Ti$_2$O$_7$ (Cd(CN)$_2$). The key difference is that the \emph{absolute} energies are about sixty times larger in Cd(CN)$_2$ than in an archetypal spin ice such as Dy$_2$Ti$_2$O$_7$.

We used these parameters to drive a series of classical Monte Carlo (MC) pseudospin simulations. Our model is subtly different to its spin-ice analogues in the sense that we have two interacting pyrochlore lattices. Nevertheless, as for the related spin-ice model \cite{Champion_2002}, we also observe an ordering transition on cooling, with $T_{\rm c}=121$\,K [Fig.~4A]. As in \cite{Champion_2002} each pyrochlore sublattice develops a nonzero magnetisation parallel to one of the cubic axes, but the two sublattice magnetisations now oppose to give a low-temperature state that is collectively antiferromagnetic. The enhancement in $T_{\rm c}$ relative to \cite{Champion_2002} indicates the dipolar interaction between lattices favours ordering. On translating pseudospins into CN$^-$ orientations, the corresponding (now antiferro\emph{electric}) state for Cd(CN)$_2$ is described by $I4_1/amd$ symmetry. So the Hamiltonian (\ref{ham2}) drives precisely the same phase transition we observe experimentally, in terms of both nature and temperature scale. Importantly, the same transition occurs for all sets of our QC interaction parameters; it is only the value of $T_{\rm c}$ that differs. Of course, longer-range interactions, strain coupling, and anharmonicity---all of which are omitted in our simple model---may mean the low-temperature $I4_1/amd$ model is not the true ground-state. In fact we may never know, since CN$^-$ reorientations are experimentally inaccessible at temperatures below 80\,K, and our different QC calculations also give a range of competing ice-rules-observing ground states whose energies differ by much less than this amount.

Just how well are other aspects of the structural behaviour of ambient-phase Cd(CN)$_2$ described by our simple spin-ice model? In Fig.~4A we show the temperature-dependent populations of CdC$_n$N$_{4-n}$ coordination environments expected as a function of temperature from our MC simulations. We find excellent consistency with the trends we determine from experimental neutron pair distribution function (PDF) and magic-angle spinning (MAS) $^{113}$Cd NMR measurements [Fig.~4B,C]. X-ray diffuse scattering also has some indirect sensitivity to CN$^-$ orientation distributions \emph{via} induced Cd displacements \cite{Fairbank_2012}. Gratifyingly, a simple Cd displacement model based on the pseudospin orientations at 298\,K is able to account for the basic form of the experimental single-crystal X-ray diffuse scattering pattern [Fig.~4D]. Our QC calculations show clearly that pseudospin fluctuations couple to significant changes (as large as $\sim0.8$\,\AA) in the accompanying Cd$\ldots$Cd distance. So even though we haven't explicitly included any spin--lattice coupling in our model, we can estimate from the temperature dependence of the MC spin projection expectation value $\langle S^2_\parallel\rangle$ the contribution of pseudospin fluctuations to thermal expansion. We determine a contribution to linear NTE of $\alpha={\rm d}(\ln a)/{\rm d}T=-6.4$\,MK$^{-1}$, which is entirely consistent with experiment ($\alpha=-20.4(4)$\,MK$^{-1}$, Ref.~\cite{Goodwin_2005}) since other mechanisms not included in our model---notably the collective CN translations of importance to NTE in Zn(CN)$_2$ \cite{Fang_2013,Collings_2013}---must also play a role.

So the key ingredients for magnetic spin-ices---namely, single-ion anisotropy and magnetic dipolar interactions---are mapped onto their electrostatic analogues in Cd(CN)$_2$ with a concomitant transformation in energy scale. The phase behaviour of this non-magnetic system---a ``dipolar structural spin-ice''---clearly obeys the same physics as the spin-ices themselves. A key experimental signature of the spin-ice state is a set of ``pinch-points'' in the magnetic diffuse scattering pattern that arises from the underlying gauge symmetry \cite{Bramwell_2011,Fennell_2014}. We cannot observe these directly for Cd(CN)$_2$ as we do not yet have access to single-crystal $^{114}$Cd(CN)$_2$ samples, but we can calculate from our MC simulations the effective magnetic diffuse scattering one would expect for an analogous spin model. Broadened pinch-point features are evident---even at room temperature [Fig.~4E]---reflecting the population of emergent monopoles \cite{Fennell_2014}. In Cd(CN)$_2$ these monopoles now correspond to C-rich or N-rich Cd coordination environments; they represent a fractionalisation of the molecular cyanide ion and would be expected to interact \emph{via} a Coulomb potential \cite{Henley_2010}. One might hope to manipulate these emergent charges by the application of an external electric field. Remarkably, the soft phonon mode responsible for the $Pn\bar3m$ to $I4_1/amd$ transition is also a manifestation of spin-ice physics \cite{Champion_2002}. This explains why ordered models do not show any lattice instabilities in DFT calculations, and points to the intriguing interplay between CN$^-$ (re)orientations---which our NMR measurements suggest take place on a timescale of a few Hz at ambient temperature---and conventional lattice dynamics that is itself reminiscent of extreme rotovibrational coupling \cite{LyndenBell_1994,Hill_2017}. Inelastic neutron scattering measurements of the dynamics of Cd(CN)$_2$ are an obvious avenue for future study. 

We finish by asking: to what extent might this behaviour be expected to generalise to other, related materials? Zn(CN)$_2$, for example, is isostructural to Cd(CN)$_2$ but does not show the same phase complexity \cite{Goodwin_2005}. Our QC calculations explain why: $J_{\rm{eff}}$ is many times smaller and $\Delta$ many times larger than for Cd(CN)$_2$---in both cases, a consequence of increased covalency and higher charge density at the Zn site. So ice-rules are not strongly enforced, nor are CN$^-$ reorientations thermally accessible. Single-network {\it s}-Cd(CN)$_2$ \cite{Phillips_2008} is likely to be more interesting since its behaviour may be describable in terms of the simpler and more strongly frustrated Dy$_2$Ti$_2$O$_7$ Hamiltonian (\ref{ham1}). If the parameters $J_{\rm{eff}}$ and $\Delta$ are comparable to those in Cd(CN)$_2$ itself, one expects the system never to order experimentally, since the phase transition temperature should be suppressed below the onset of orientational glass formation. This is consistent with the absence of any phase transition in variable-temperature (100--300\,K) X-ray diffraction measurements \cite{Phillips_2008}. Indeed the shift in temperature scale may provide new avenues for exploring metastability in a spin-ice-like system \cite{Giblin_2018}. Cd(CN)$_2$ also forms a very large array of host--guest structures, many of which are based on the pyrochlore lattice \cite{Kitazawa_1994}. This opens up the unexpected possibility of tuning spin-ice behaviour \emph{via} guest (de)sorption. The substitution of CN$^-$ for Br$^-$---as explored historically in the context of alkali cyanide quadrupolar spin-glass analogues \cite{Binder_1998}---will have an effect equivalent to doping a spin-ice with nonmagnetic impurities. Likewise, pressure is an as-yet unexplored variable for spin-ice physics that is now suddenly accessible given the shift in temperature scale. One way or the other, our study has reinforced the concept that materials with strongly-correlated structural disorder can mirror the remarkable physics of exotic electronic phases \cite{Keen_2015,Cairns_2016}. But it demonstrates also how the theory that underpins our understanding of the latter helps rationalise the phase behaviour of the former. Noting the empirical mapping between symmetry breaking in Cd(CN)$_2$ and the VII/VIII proton-ordering transition in water ice \cite{Kuo_2004}, for example, one might reasonably ask whether the phenomenology of spin-ices may yet shed light on the physics of their fundamentally important parent: water ice itself.

\cleardoublepage

\clearpage

\section*{Acknowledgements}
The authors gratefully acknowledge funding from the E.P.S.R.C. (Grants  EP/G004528/2, EP/L000202, EP/R029431), the E.R.C. (Grant 788144), the Leverhulme Trust (Grant RPG-2015-292), the Swiss National Science Foundation (Fellowship to A. S.), and St John's College, Oxford (Fellowship to J. W. M.). Single crystal X-ray diffuse scattering measurements were performed on beamline BM01 at the European Synchrotron Radiation Facility (ESRF), Grenoble, France. We are grateful to Dmitry Chernyshov (ESRF) for providing assistance in using the beamline. We gratefully acknowledge useful discussions with Lucy Clark (Liverpool) and Joseph Paddison (Cambridge).

\clearpage

\noindent {\bf Fig.\ 1.} \\
{\bf Ice rules on the pyrochlore lattice.} (A) Hydrogen bond orientations in water ice and (B) magnetic moment orientations in rare-earth spin-ices both obey the same ``two-in-two-out'' rule for each tetrahedral unit of the pyrochlore lattice (black lines). The same rules are thought to apply to cyanide ion orientations in Cd(CN)$_2$, the crystal structure of which is represented in (C). Cd atoms shown as green spheres and CN$^-$ ions as ellipsoids; the unit cell (outlined in red) corresponds to one octant of the underlying pyrochlore lattice (outlined in black). For clarity, only one of the two interpenetrating pyrochlore lattices is shown. In the average structure of Cd(CN)$_2$, CN$^-$ orientations are disordered (grey). This disorder is not random: there is a preference for each Cd to bind two C atoms (white hemiellipsoids) and two N atoms (black hemiellipsoids) in an icelike ``two-in-two-out'' arrangement.

\clearpage

\noindent {\bf Fig.\ 2.} \\
{\bf Low-temperature crystallography of Cd(CN)$_2$.} (A) Intensity map of the temperature-dependent neutron powder diffraction pattern of Cd(CN)$_2$, showing the existence of a phase transition at $T_{\rm c}=130$\,K. (B) Rietveld fits to the diffraction pattern at representative temperatures above (top, $Pn\bar3m$) and below (bottom, $I4_1/amd$) $T_{\rm c}$. Data are shown in black, fits in red, difference in grey and reflection positions as blue vertical bars. The contribution from a minor Hg(CN)$_2$ impurity \cite{Coates_2018} is indicated by asterisks. (C) Temperature evolution of the spontaneous strain (red circles; extracted from the renormalised lattice parameters, inset) and long-range CN$^-$ orientational order (black circles) determined by Rietveld refinement. Note the onset of the CN$^-$ reorientational glass transition at $T_{\rm g}\simeq 80$\,K affects the latter more obviously than the former. (D) Representation of the $I4_1/amd$ crystal structure of Cd(CN)$_2$ at 10\,K: Cd atoms are shown in green, C in white, and N in black. Thermal ellipsoids (isotropic) are shown at 50\% probability.

\clearpage

\noindent {\bf Fig.\ 3.} \\
{\bf Microscopic single-ion and pairwise interaction parameters in Cd(CN)$_2$.} (A) The effective exchange interaction (strength $J_{\rm eff}$) operates between nearest-neighbours within the same pyrochlore sublattice. Dipolar interactions (strength $D$) give rise to an effective exchange interaction between nearest neighbours on alternate sublattices. The single-ion anisotropy term $\Delta$ reflects the enthalpy barrier to CN$^-$ reorientations. (B) Relative DFT energies $E_{\rm rel}$ for a series of single network (blue symbols) and interpenetrated (red symbols; shifted vertically by 1\,kJ\,mol$^{-1}$) Cd(CN)$_2$ configurations with different CdC$_n$N$_{4-n}$ coordination environments: CdC$_2$N$_2$ (geometric parameter $\gamma = \frac{2}{3}(n-2)^2=0$), CdC$_3$N/CdCN$_3$ ($\gamma=\frac{2}{3}$), and CdC$_4$/CdN$_4$ ($\gamma=\frac{8}{3}$). Simple geometric arguments give $E_{\rm rel}(\gamma)=\gamma J_{\rm eff}$ and $\gamma(J_{\rm eff}+D)$ for single-network and interpenetrated Cd(CN)$_2$, respectively. Shaded regions highlight the range of linear fits obtained to these data. (C) Values of $J_{\rm eff}$ (blue) and $D$ (red) extracted from the slopes of the linear fits to the data in (B) for a variety of density functionals. The NMR-derived $J_{\rm eff}$ and geometric $D$ values are shown as horizontal lines or bands. (D) Representative nudged elastic band calculation energies (PBE-D3 functional) for a 180$^\circ$ CN$^-$ orientation flip (filled green symbols) and corresponding fit $E_{\rm rel}(\theta)=\Delta\cos^2\theta=\Delta S_{\parallel}^2$ (solid green line), from which the value of $\Delta$ was obtained. The $\theta\simeq90^\circ$ transition state involves a C-bridged Cd--(CN)--Cd linkage (inset) and a reduced Cd$\ldots$Cd separation.

\clearpage

\noindent {\bf Fig.\ 4.} \\
{\bf Calculated and observed structural spin-ice behaviour in Cd(CN)$_2$.} (A) Temperature dependent population of CdC$_n$N$_{4-n}$ coordination environments as determined by our MC pseudospin simulations (small symbols), neutron PDF measurements (open squares), and $^{113}$Cd MAS NMR spectroscopic measurements (open circles), compared with the non-interacting analytical result (solid lines, Ref.~\citenum{Nishikiori_1990}). The sublattice magnetisation (red symbols, red line)---or polarisation, in the specific context of Cd(CN)$_2$---is an order parameter for the transition between the disordered ($Pn\bar3m$) spin-ice state and the low-temperature ($I4_1/amd$) antiferroelectric state, which we find to occur at 121\,K. (B) Variable-temperature $^{113}$Cd MAS NMR spectra (black lines) and corresponding fits (red lines) used to extract the values given in (A). (C) Variable-temperature neutron PDF data (lower curves) and calculated PDFs from $\gamma=0$ and $\frac{2}{3}$ QC configurations (upper curves) used to determine the PDF-derived values in (A). (D) Single-crystal X-ray diffuse scattering pattern ($(hk0)$ plane) measured at 298\,K and calculated from a coupled CN-orientation/Cd-displacement model based on the 298\,K pseudospin configuration. (E) Effective magnetic diffuse scattering pattern ($(hhl)$ plane) extracted from our 300\,K MC simulations by interpreting CN$^-$ orientations from a single pyrochlore sublattice as classical spin vectors with the Dy$^{3+}$ magnetic form factor, compared with the experimental magnetic diffuse scattering pattern measured for spin-ice Dy$_2$Ti$_2$O$_7$ at 1.3\,K \cite{Fennell_2004}. The persistence of broadened pinch-points in the former---characteristic of the population of monopole defects within a structural-spin-ice state---is evident in the ($\sim$ constant) longitudinal and (peaked) transverse intensity cuts (blue and red lines; see inset).

\clearpage

\begin{center}
\includegraphics{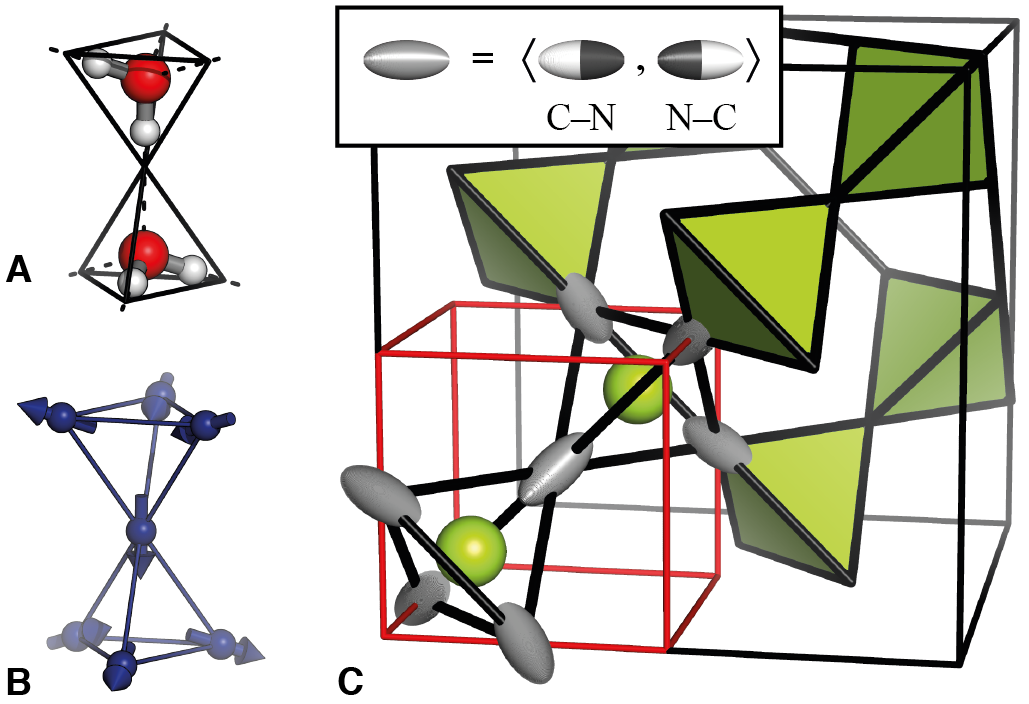}\\
FIGURE 1\\
\end{center}

\clearpage

\begin{center}
\includegraphics{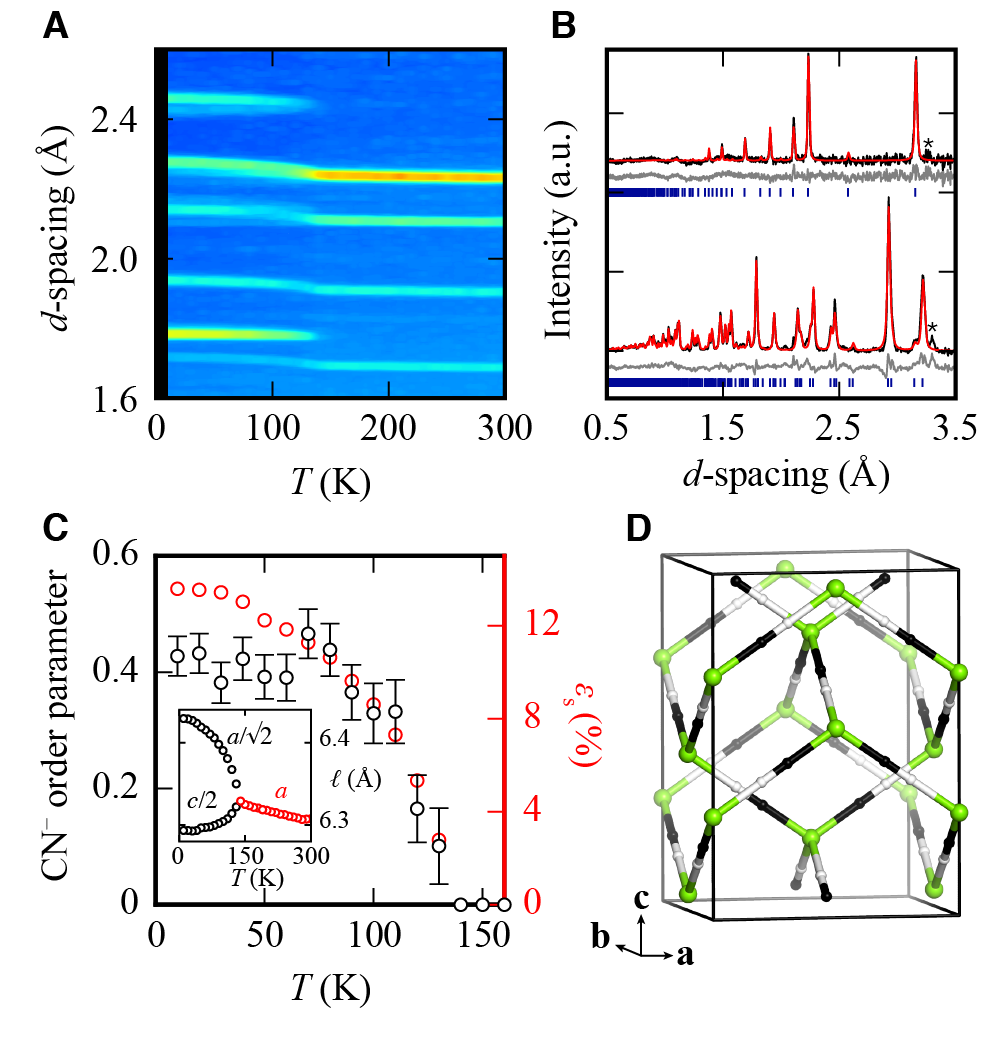}\\
FIGURE 2\\
\end{center}

\clearpage

\begin{center}
\includegraphics{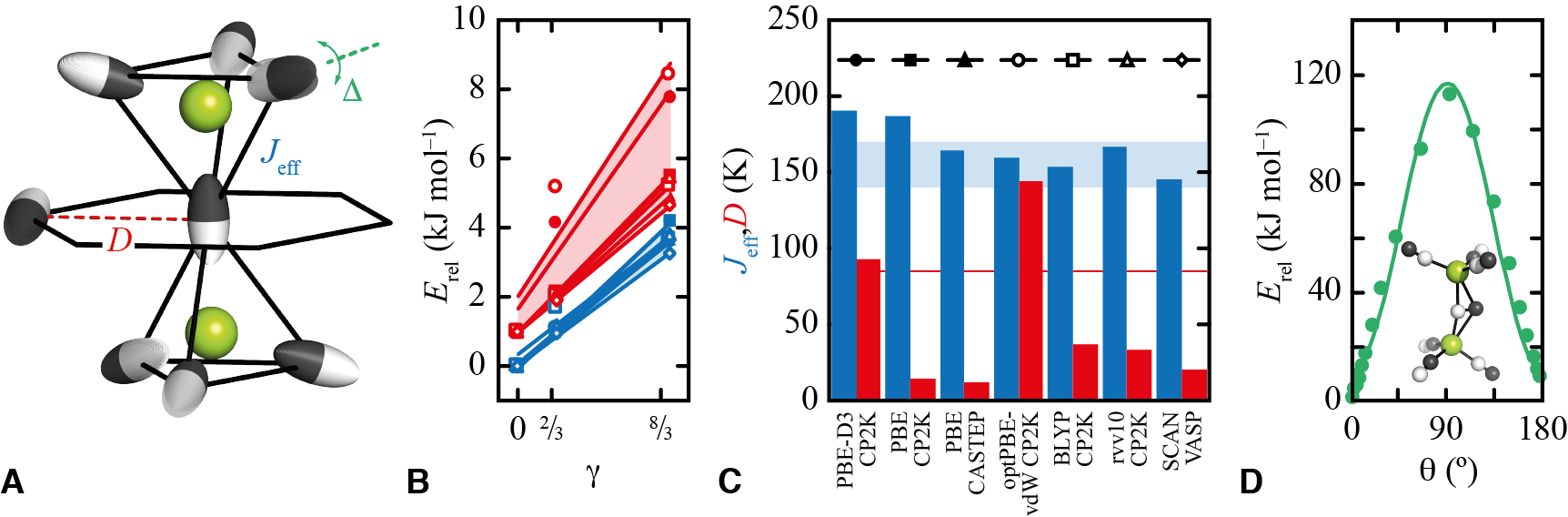}\\
FIGURE 3\\
\end{center}

\clearpage

\begin{center}
\includegraphics{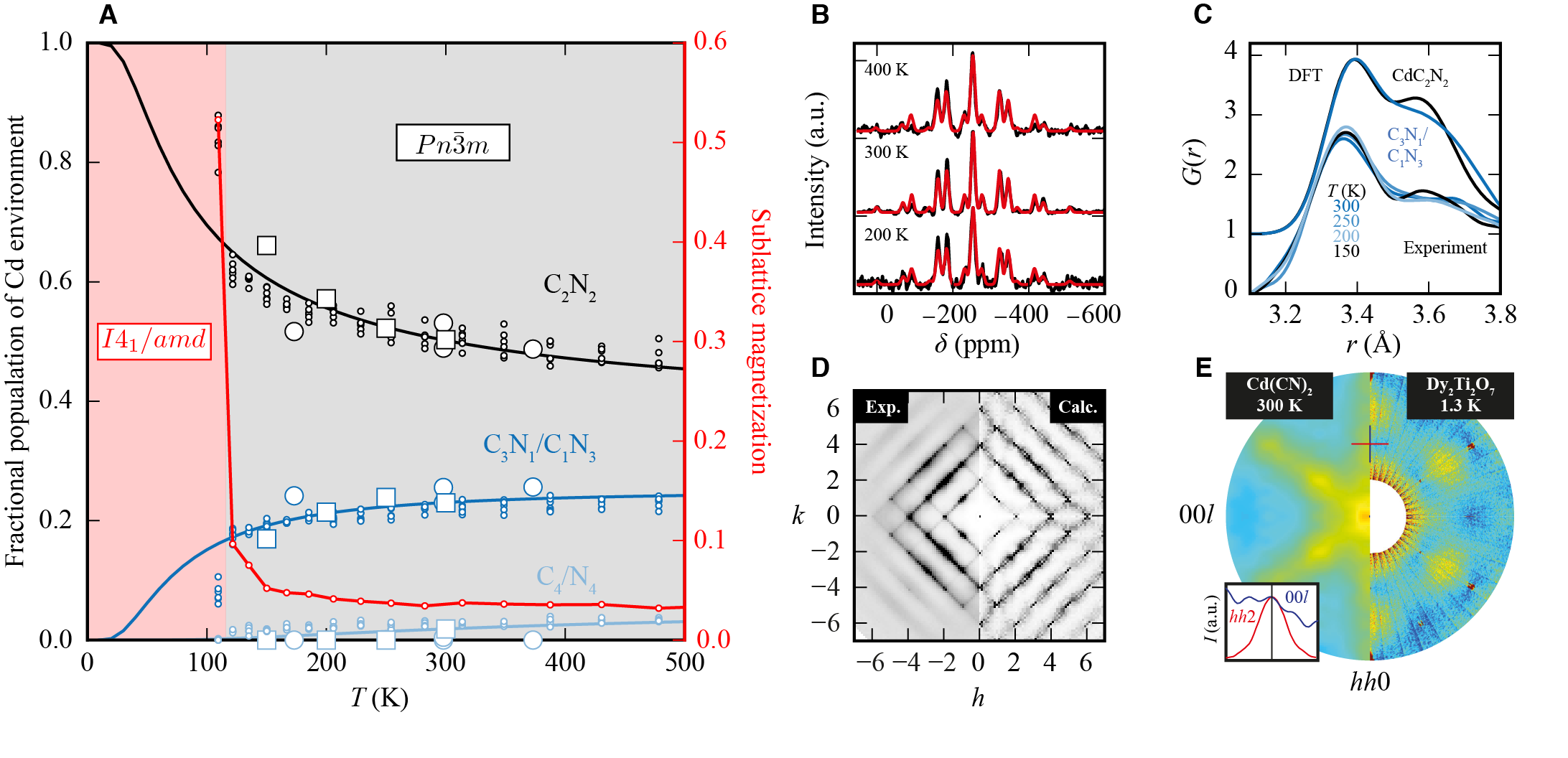}\\
FIGURE 4\\
\end{center}

\clearpage


\begin{thebibliography}{10}
\bibitem{notenotcaseforxrays}
This is not the case for X-ray scattering measurements (both in-house and synchrotron), where we observe a variety of complex additional transitions that are frequently irreversible and arise as a consequence of X-ray beam damage.
\bibitem{noterefactor}
Note that the mapping of anisotropic Heisenberg spins to Ising variables introduces a factor of $\frac{1}{3}$, which we have taken into account in our comparison here.
\end{thebibliography}
\end{document}